\documentclass[runningheads]{llncs}
\usepackage[T1]{fontenc}
\usepackage{graphicx}
\usepackage{booktabs}
\usepackage{microtype}

\begin{document}
\title{Is Deformable Image Registration Ready for Brain Metastasis Reirradiation Dose Accumulation? A Longitudinal MRI Benchmark of Registration Accuracy}
\titlerunning{Benchmarking DIR for Brain Metastasis Reirradiation}
%
\author{
Hengjie Liu\inst{1} \and
Manju Sharma\inst{1} \and
Xinyi Fu\inst{1} \and
Di Xu\inst{1} \and
Ke Sheng\inst{1}
}

\authorrunning{H. Liu et al.}
%
\institute{
Department of Radiation Oncology, University of California, San Francisco,
San Francisco, CA, USA
}

\maketitle              

\begin{abstract}
Dose accumulation is increasingly important in adaptive radiation therapy and reirradiation, but its clinical validity depends on the performance of deformable image registration (DIR). Reirradiation of brain metastases (BMs) with stereotactic radiosurgery (SRS) provides a controlled but clinically meaningful DIR test case: intra-subject brain deformation is usually limited after rigid alignment, yet recurrent lesions can undergo substantial local shape and volume changes that rigid registration cannot capture and can affect dose accumulation. We benchmarked a wide range of learning-based and optimization-based DIR methods on 87 manually screened longitudinal contrast-enhanced T1-weighted MRI lesion pairs from an institutional BM SRS retreatment cohort. Learning-based methods pretrained on healthy-brain MRI were evaluated zero-shot and after instance-specific optimization (ISO) or tumor-proximity target-specific optimization (TSO). Registration was assessed using lesion overlap (Dice), surface distance metrics (HD95 and sASD), target-volume recovery, and runtime and memory. Pretrained learning-based methods showed variable zero-shot performance, while ISO/TSO improved all tested learning-based families. However, optimization-based methods remained the best-performing approach while maintaining reasonable runtime. These findings suggest that even state-of-the-art DIR methods do not yet provide sufficiently accurate and consistent registration for unmonitored use in brain metastasis reirradiation dose accumulation. Because accurate registration is a prerequisite for deformable dose accumulation, clinical application will require case-level quality control and direct assessment of how registration uncertainty affects downstream dose metrics.

\keywords{Deformable image registration \and Dose accumulation \and Reirradiation \and Brain metastases \and Stereotactic radiosurgery.}
\end{abstract}

\section{Introduction}
Deformable image registration (DIR) has several important applications in radiation therapy (RT), including contour propagation, multimodal image fusion, adaptive planning, and dose accumulation~\cite{Brock2017TG132}. Its role is especially important when anatomy and delivered dose must be compared across time, as in adaptive RT and reirradiation. However, clinical adoption of DIR remains challenging. The problem is ill-posed, voxel-level ground truth is rarely available, and registration errors can propagate into accumulated dose estimates~\cite{Sotiras2013DIRSurvey,Nenoff2023DIRUncertainty,Meyer2025DIRUncertaintyDose}. Registration performance and uncertainty must therefore be carefully evaluated in clinically relevant settings, together with their effects on downstream analyses.

Reirradiation of brain metastases (BMs) with stereotactic radiosurgery (SRS) is one such setting. BMs are common in patients with systemic cancer, and SRS, including Gamma Knife treatment, is a standard focal treatment for selected patients. As systemic and local therapies improve, more patients survive long enough to develop local recurrence after prior SRS, making repeat focal treatment increasingly relevant~\cite{Loi2020StereotacticReirradiation,Kowalchuk2022ReirradiationSRS}. In these patients, tumor control must be balanced against adverse radiation effect, preserving neurologic function and quality of life. Previous SRS studies of BMs have linked toxicity risk to dose-volume measures such as V10, V12, and irradiated volume, while repeat-SRS studies have shown that reirradiation is feasible but clinically heterogeneous~\cite{Milano2021HyTECBrain,Minniti2011BrainRN,Blonigen2010IrradiatedVolume,Loi2020StereotacticReirradiation,Kowalchuk2022ReirradiationSRS}. Estimating the combined dose from multiple treatments could therefore support clinical analysis, but the validity of those estimates depends directly on registration quality.

Recent developments in DIR have increasingly focused on learning-based methods, which offer fast inference and have achieved strong performance on established benchmarks~\cite{Hering2023Learn2Reg,Chen2025DLSurveyRegistration}. Brain MRI has been a major focus, supported by large public datasets and registration challenges such as LUMIR, a large-scale inter-subject brain MRI registration benchmark comprising more than 3,000 healthy-brain scans~\cite{Chen2026LUMIR}. However, strong benchmark performance does not necessarily translate to real clinical data. Models trained mainly on healthy-brain MRI may not generalize reliably to scans containing contrast-enhancing tumors, treatment-related changes, and substantial local lesion evolution. Recent studies have also shown that learning-based registration can be sensitive to preprocessing, evaluation protocols, and domain shift, with performance decreasing on out-of-distribution data~\cite{Jena2024Mirage,Jena2026LUMirage}. Instance-specific optimization can help adapt pretrained models to individual image pairs~\cite{Mok2023NIO}. Optimization-based methods remain valuable alternatives and important practical baselines, particularly when accuracy and robustness are more important than immediate inference speed~\cite{Jena2026FireANTs,Jena2026LUMirage}.

In this study, we benchmark representative learning-based and optimization-based DIR methods on longitudinal contrast-enhanced T1-weighted MRI from a cohort of patients undergoing repeat SRS for BMs. Using publicly available models pretrained for LUMIR, we evaluate learning-based registration under zero-shot inference and after instance-specific optimization (ISO) or tumor-proximity target-specific optimization (TSO), and compare the results with conventional and modern optimization-based methods. Registration performance is assessed using lesion overlap, surface distance, target-volume recovery, runtime, and memory. Our results show that models pretrained on healthy-brain MRI do not consistently perform well out of the box, whereas ISO and TSO improve all tested learning-based method families. Nevertheless, optimization-based methods remain the most accurate and robust overall, with modern GPU-optimized implementations achieving reasonable runtimes. The substantial variation across methods and cases suggests that DIR is not yet ready to be used as an automatic and unchecked step in dose accumulation for BM reirradiation. Further methodological improvement, registration quality screening, and sensitivity analysis of downstream clinical dose metrics are needed before accumulated dose estimates can be interpreted with confidence.

\section{Methods}
\subsection{Study Cohort and Registration Task}
This study used longitudinal contrast-enhanced T1-weighted MRI from patients with brain metastases treated with repeat stereotactic radiosurgery (SRS). Each lesion pair consisted of a recurrent lesion with MRI available at both the first and second SRS time points. Following rigid alignment, the earlier MRI was warped into the coordinate system of the later MRI to support subsequent deformable dose mapping and summation. Clinician-delineated enhancing tumor contours were available on the contrast-enhanced T1-weighted MRI at both SRS time points and served as the reference lesion masks for lesion-pair screening and registration evaluation.

\begin{figure}
\centering
\includegraphics[width=0.80\textwidth]{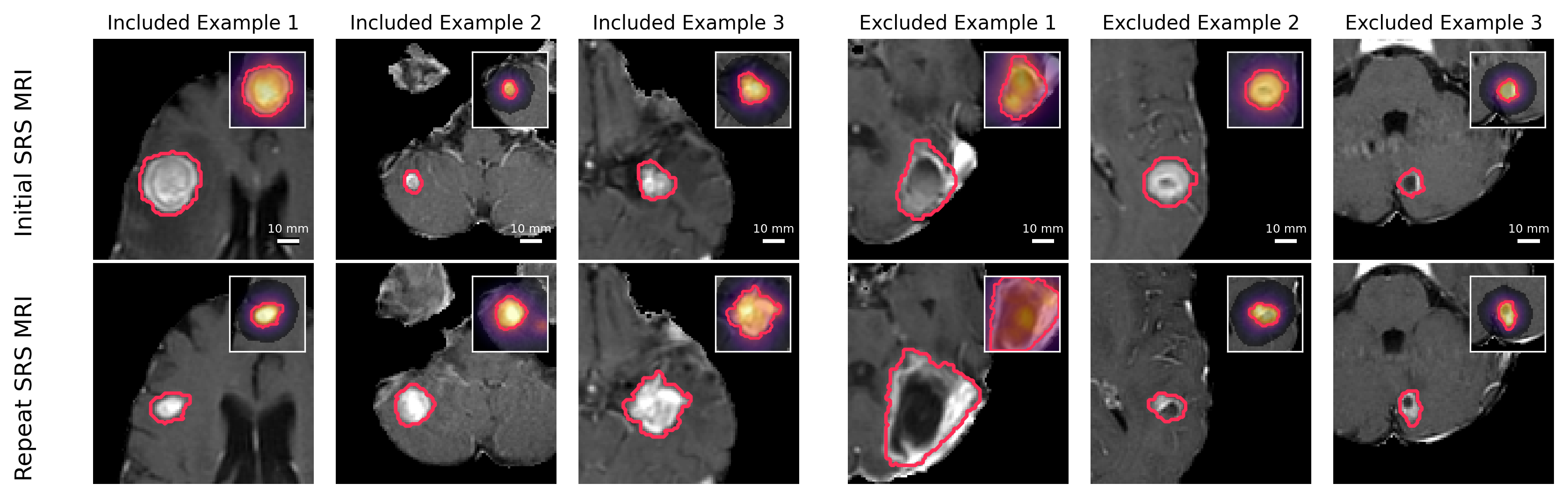}
\caption{Representative included and excluded lesion pairs. Red contours show lesion masks on initial and repeat SRS MRI; insets show local dose overlays. Excluded cases had poor intensity correspondence for DIR.}
\label{fig:example}
\end{figure}

Because the benchmark was designed to evaluate intensity-based deformable image registration (DIR), lesion pairs were manually screened for suitability. Included pairs required contrast-enhanced T1-weighted MRI at both time points and sufficient similarity in normal brain anatomy and lesion enhancement to support intensity-based matching. Pairs were excluded when the recurrent lesion developed major new internal structures, such as a large necrotic or cystic cavity; when corresponding enhancing structures were absent or substantially altered between time points; or when adjacent enhancing vessels made the lesion boundary ambiguous. From an initial institutional cohort of 97 patients with 171 recurrent lesions, 55 patients with 87 lesion pairs were retained for the DIR benchmark. Representative included and excluded lesion pairs are shown in Fig.~\ref{fig:example}.

\subsection{Deformable Registration Formulation}
Let $M$ denote the moving image at the earlier SRS time point and $F$ denote the fixed image at the later SRS. DIR estimates a deformation $\phi_u$, parameterized by a displacement field $u$, that maps the moving image into the fixed image space:
\begin{equation}
\phi_u(x) = x + u(x).
\label{eq:deformation}
\end{equation}

For each image pair, optimization-based registration solves a pair-specific objective $\mathcal{L}$:
\begin{equation}
u^{*}
= \mathop{\mathrm{arg\,min}}_{u} \mathcal{L}(F,M;u)
= \mathop{\mathrm{arg\,min}}_{u}
\mathcal{D}\bigl(F, M \circ \phi_u\bigr) + \lambda \mathcal{R}(u).
\label{eq:pairwise_objective}
\end{equation}
Here, $\mathcal{D}$ measures image dissimilarity, $\mathcal{R}$ regularizes the deformation field, and $\lambda$ controls the tradeoff between image matching and deformation smoothness~\cite{Sotiras2013DIRSurvey}.

Learning-based registration uses the same loss principle, but learns a network $g_{\theta}$ that predicts the displacement field from an image pair. Training minimizes the expected registration loss over a training distribution:
\begin{equation}
\theta^{*}
= \mathop{\mathrm{arg\,min}}_{\theta}
\mathrm{E}_{(F_i,M_i)\sim \mathcal{P}}
\mathcal{L}\bigl(F_i,M_i;g_{\theta}(F_i,M_i)\bigr).
\label{eq:learning_objective}
\end{equation}
At inference, pretrained registration uses $u_{\mathrm{PT}}=g_{\theta^{*}}(F,M)$. This is computationally efficient, but depends on how well the test images match the training distribution. We therefore followed the preprocessing used by the LUMIR-style models, including skull stripping~\cite{Hoopes2022SynthStrip}, template alignment, resampling to isotropic $1~\mathrm{mm}$, and intensity normalization. The image pair was rigidly aligned using the robust rigid registration method in FreeSurfer~\cite{Reuter2010RobustRegister}. Even with matched preprocessing, models pretrained on healthy-brain inter-subject MRI may remain suboptimal for longitudinal intra-subject contrast-enhanced MRI with recurrent tumors. This motivates pair-specific adaptation.

\subsection{Instance-Specific and Target-Weighted Optimization}
\label{subsec:iso_tso}
For learning-based methods, we evaluated three modes: PT, ISO and TSO. Pretrained inference (PT) used the released model directly without adaptation. Instance-specific optimization (ISO) initialized the network with its pretrained parameters and fine-tuned the network parameters on each individual test image pair by minimizing the registration loss.
\begin{equation}
\theta_{\mathrm{ISO}}^{*}
= \mathop{\mathrm{arg\,min}}_{\theta}
\mathcal{L}\bigl(F,M;g_{\theta}(F,M)\bigr),
\qquad
\theta_0 = \theta_{\mathrm{PT}},
\qquad
u_{\mathrm{ISO}}^{*}=g_{\theta_{\mathrm{ISO}}^{*}}(F,M).
\label{eq:iso}
\end{equation}

Because this benchmark focuses on recurrent lesions, we also evaluated tumor-proximity target-specific optimization (TSO). TSO used a spatial weight map to emphasize image matching near the lesion while retaining a nonzero contribution from the rest of the brain. Let $T$ denote the target lesion region and $d_T(x)$ the distance from voxel $x$ to $T$:
\begin{equation}
w_T(x) = w_{\min} + (1-w_{\min})\exp\left(-\frac{d_T(x)}{h}\right).
\label{eq:tso_weight}
\end{equation}
For learning-based methods, TSO used the same 100-step pair-specific optimization as ISO, but applied this target-proximity weighting to the loss function.

The same spatial weighting strategy was also applied to optimization-based registration using FireANTs, whose GPU-accelerated PyTorch implementation supports flexible customization of the registration objective. These target-weighted variants were optimized independently for each image pair and are denoted by the suffix ``-TW.''

\subsection{Registration Methods}
The learning-based methods included VoxelMorph (VM), TransMorph (TM), vector field attention (VFA), SITReg (SIT), and UniGradICON (UGI). VM and TM were included as widely used baseline architectures~\cite{Balakrishnan2019VoxelMorph,Chen2022TransMorph}. VFA and SIT were included as recent high-performing brain MRI registration models evaluated in the LUMIR setting~\cite{Liu2024VFA,Honkamaa2024SITReg,Chen2026LUMIR}. UGI was included as a foundation-style registration model intended for broad zero-shot use~\cite{Tian2024UniGradICON}.

Optimization-based methods were included as training-free pairwise baselines. ANTs-SyN and Greedy represented established conventional registration methods~\cite{Avants2008ANTsSyN,Yushkevich2016Greedy}. We also evaluated the modern GPU-based methods FireANTs and SINR. FireANTs was included as a GPU-accelerated, multiscale diffeomorphic registration method~\cite{Jena2026FireANTs}. We evaluated both FireANTs Greedy and FireANTs SyN, abbreviated FA-Greedy and FA-SyN, together with their target-weighted variants, FA-Greedy-TW and FA-SyN-TW, as defined in Sect.~\ref{subsec:iso_tso}. Finally, we evaluated SINR, a spline-enhanced implicit neural representation (INR) approach that parameterizes the deformation field using a pair-specific multi-layer perceptron (MLP)~\cite{SideriLampretsa2024SINR}.

\subsection{Implementation Details}
All learning-based methods used their publicly released pretrained checkpoints. For ISO and TSO, the network parameters were optimized independently for each image pair for 100 steps using Adam with a learning rate of $10^{-4}$. The original image-similarity and deformation-regularization terms were retained. The 100-step schedule was selected based on preliminary convergence analysis, which showed that the registration loss had stabilized by this point. TSO constructed the spatial weight map from the lesion contour on the later MRI using a 5-mm expansion, with $w_{\min}=0.1$ and $h=5~\mathrm{mm}$. ANTs-SyN and Greedy followed the high-performing configurations for brain MRI reported by Jena et al.~\cite{Jena2024Mirage}. SINR used its default configuration with the control-point spacing set to 4 voxels. FireANTs hyperparameters were selected on a subset of cases and then fixed for evaluation, using a four-level image pyramid and an NCC window size of 21. Experiments were performed using a single NVIDIA RTX A6000 GPU. Reported runtimes include registration only and exclude common preprocessing.

\begin{figure}
\centering
\includegraphics[width=\textwidth]{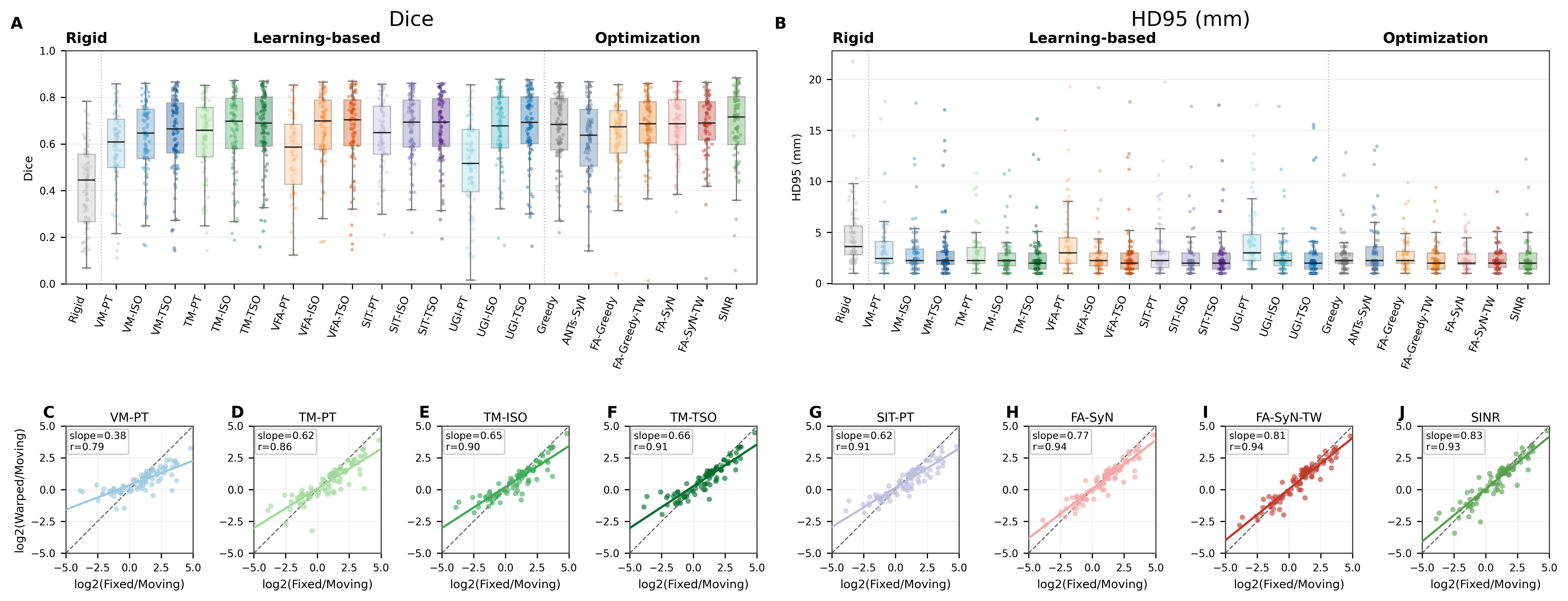}
\caption{Registration accuracy and target-volume recovery. (A-B) Dice and HD95 for rigid, learning-based, and optimization-based methods. (C-J) Recovered versus observed log2 target-volume change for selected methods; dashed lines indicate ideal recovery.}
\label{fig:results}
\end{figure}

\subsection{Evaluation Metrics}
Registration accuracy was evaluated by warping the earlier lesion mask into the later MRI space and comparing it with the later lesion mask. We report Dice, 95th percentile Hausdorff distance (HD95), and symmetric average surface distance (sASD). Dice measures volumetric overlap, while HD95 and sASD measure boundary agreement.

To assess target-volume recovery, we computed the moving lesion volume $V_M$, fixed lesion volume $V_F$, and warped moving lesion volume $V_W$. Observed volume change was defined as $\log_{2}(V_F/V_M)$, and recovered volume change as $\log_{2}(V_W/V_M)$. Perfect recovery follows the line $y=x$. For each method, we report the regression slope and Pearson correlation between recovered and observed log2 volume change. A slope near 1 with high correlation indicates better recovery of target-volume change; slopes below 1 indicate under-recovery of volume change. Runtime and peak memory were also recorded for each method.

\begin{table}[t]
\caption{Registration accuracy and target-volume agreement on the 87-case cohort. Dice, HD95, and sASD are reported as mean (standard deviation). HD95 and sASD are in mm. Volume agreement reports the regression slope and Pearson correlation between warped and fixed target volumes. Bold indicates the top three methods for each metric; HD95/sASD are ranked lower-is-better, slope by closeness to 1, and $r$ higher-is-better.}
\label{tab:registration_results}
\centering
\footnotesize
\setlength{\tabcolsep}{2.6pt}
\renewcommand{\arraystretch}{0.88}
\begin{tabular}{lccccc}
\toprule
Method & Dice & HD95 & sASD & Vol. slope & Vol. $r$ \\
\midrule
\multicolumn{6}{l}{\textit{Baseline}} \\
Rigid & 0.417 (0.173) & 4.62 (3.10) & 2.23 (1.53) & -- & -- \\
\midrule
\multicolumn{6}{l}{\textit{Learning-based}} \\
VM-PT & 0.582 (0.172) & 3.44 (2.81) & 1.03 (0.81) & 0.384 & 0.786 \\
VM-ISO & 0.622 (0.163) & 3.07 (2.54) & 0.88 (0.68) & 0.515 & 0.873 \\
VM-TSO & 0.640 (0.174) & 3.02 (2.75) & 0.83 (0.71) & 0.560 & 0.866 \\
\addlinespace[2pt]
TM-PT & 0.629 (0.166) & 2.92 (1.81) & 0.81 (0.45) & 0.623 & 0.857 \\
TM-ISO & 0.656 (0.169) & 2.64 (1.85) & 0.73 (0.44) & 0.645 & 0.900 \\
TM-TSO & 0.667 (0.164) & 2.73 (2.40) & 0.74 (0.59) & 0.658 & 0.908 \\
\addlinespace[2pt]
VFA-PT & 0.552 (0.186) & 3.92 (3.23) & 1.24 (1.11) & 0.328 & 0.625 \\
VFA-ISO & 0.661 (0.159) & 2.73 (2.39) & 0.76 (0.64) & 0.648 & 0.887 \\
VFA-TSO & 0.661 (0.170) & 2.80 (2.65) & 0.77 (0.70) & 0.642 & 0.877 \\
\addlinespace[2pt]
SIT-PT & 0.642 (0.145) & 3.15 (2.87) & 0.84 (0.68) & 0.617 & 0.912 \\
SIT-ISO & 0.669 (0.143) & 2.68 (2.31) & \textbf{0.71 (0.51)} & 0.696 & 0.923 \\
SIT-TSO & 0.669 (0.148) & 2.69 (2.40) & \textbf{0.72 (0.55)} & 0.705 & 0.919 \\
\addlinespace[2pt]
UGI-PT & 0.515 (0.186) & 4.25 (3.18) & 1.32 (1.09) & 0.298 & 0.716 \\
UGI-ISO & 0.664 (0.159) & 2.75 (2.30) & 0.76 (0.58) & 0.646 & 0.891 \\
UGI-TSO & 0.674 (0.159) & 2.73 (2.65) & 0.73 (0.63) & 0.678 & 0.911 \\
\midrule
\multicolumn{6}{l}{\textit{Optimization-based}} \\
Greedy & 0.664 (0.155) & 2.63 (1.81) & 1.00 (0.64) & 0.663 & 0.900 \\
ANTs-SyN & 0.624 (0.158) & 3.15 (2.46) & 1.21 (0.88) & 0.552 & 0.868 \\
FA-Greedy & 0.642 (0.145) & 2.81 (1.62) & 1.08 (0.55) & 0.669 & 0.929 \\
FA-Greedy-TW & \textbf{0.676 (0.140)} & 2.51 (1.46) & 0.96 (0.46) & 0.755 & \textbf{0.943} \\
FA-SyN & 0.675 (0.136) & \textbf{2.41 (1.20)} & 0.92 (0.32) & \textbf{0.772} & \textbf{0.940} \\
FA-SyN-TW & \textbf{0.681 (0.139)} & \textbf{2.38 (1.22)} & 0.91 (0.32) & \textbf{0.805} & \textbf{0.944} \\
SINR & \textbf{0.680 (0.160)} & \textbf{2.45 (1.66)} & \textbf{0.66 (0.36)} & \textbf{0.825} & 0.925 \\
\bottomrule
\end{tabular}
\end{table}

\section{Results and Discussion}

Registration results are summarized in Fig.~\ref{fig:results} and Table~\ref{tab:registration_results}. Runtime and memory are summarized in Table~\ref{tab:runtime_memory}.

Pretrained learning-based methods showed strongly model-dependent zero-shot performance. Among the pretrained models, SIT performed best, consistent with the LUMIR challenge results~\cite{Chen2026LUMIR}. In contrast, UGI, despite being proposed as a foundation-style registration model, did not generalize well to this longitudinal BM reirradiation cohort compared with models trained specifically for healthy inter-subject brain MRI registration.

Both ISO and TSO improved over pretrained inference, supporting pair-specific adaptation as a strategy for reducing the domain gap between the pretraining task of inter-subject healthy-brain MRI registration and the target task of longitudinal contrast-enhanced brain MRI registration with recurrent enhancing tumors. The largest gains were observed for models with weaker zero-shot performance, including VFA and UGI. TSO did not uniformly improve beyond ISO. For methods with explicit multi-resolution coarse-to-fine optimization, such as VFA and SIT, ISO appeared to capture most of the benefit, and adding tumor-proximity weighting provided little additional improvement. This highlights the effectiveness of explicit multi-resolution design to simultaneously capture global and local deformation~\cite{RethinkReg2024,MUSA2025,RevisitDIR2025}.

Optimization-based methods remained the strongest overall. Modern methods, particularly FireANTs-SyN and SINR, achieved the best registration accuracy. Target weighting further improved FireANTs, especially for the Greedy variant, showing that lesion-focused weighting can be similarly helpful as in learning-based methods.

Target-volume recovery followed the same general pattern. Pretrained learning-based methods often under-recovered observed lesion volume change, while ISO/TSO improved volume agreement. The strongest volume-recovery trends were seen for modern optimization-based methods, especially SINR and target-weighted FireANTs.

Runtime and memory showed a practical tradeoff. Pretrained learning-based inference was fastest, but ISO/TSO substantially increased runtime and GPU memory use. FireANTs achieved strong accuracy with relatively modest runtime and memory demand, making modern optimization-based registration the most promising near-term approach for dose-accumulation workflows for BM reirradiation.

\begin{table}[t]
\caption{Runtime and peak GPU memory. Values are reported as mean runtime and peak memory. CPU-only methods do not report GPU memory.}
\label{tab:runtime_memory}
\centering
\footnotesize
\setlength{\tabcolsep}{4.0pt}
\renewcommand{\arraystretch}{0.90}
\begin{tabular}{lcc}
\toprule
Method & PT & ISO/TSO or optimization \\
\midrule
VoxelMorph & 0.23 s / 2.9 GB & 94 s / 13 GB \\
TransMorph & 0.30 s / 1.4 GB & 124 s / 9.4 GB \\
VFA & 0.69 s / 8.7 GB & 186 s / 40 GB \\
SITReg & 1.33 s / 4.9 GB & 410 s / 28 GB \\
UniGradICON & 0.62 s / 2.7 GB & 203 s / 12 GB \\
\midrule
Greedy (CPU) & -- & 344 s / -- \\
ANTs SyN (CPU) & -- & 316 s / -- \\
FireANTs Greedy & -- & \textbf{13 s} / \textbf{2.6 GB} \\
FireANTs SyN & -- & \textbf{20 s} / \textbf{4.2 GB} \\
SINR cps4 & -- & 159 s / \textbf{3.7 GB} \\
\bottomrule
\end{tabular}
\end{table}

\section{Conclusion}
This work establishes a foundation for evaluating deformable dose accumulation for brain metastasis reirradiation. Our benchmark demonstrates that DIR for longitudinal BM reirradiation remains an important topic of investigation. Although pair-specific adaptation consistently improves pretrained learning-based models, modern optimization-based methods, particularly FireANTs and SINR, provide the strongest overall geometric performance while maintaining practical runtimes. Nevertheless, performance remains variable at the case level, and even state-of-the-art DIR methods do not yet provide sufficiently accurate and consistent registration for unmonitored use in dose-accumulation workflows.

The present study only evaluates lesion correspondence. Future work will extend this framework to downstream dose and outcome analyses by developing explicit case-level quality-control and registration-acceptance criteria, improving or excluding registrations that do not meet these criteria, and restricting dose analysis to quality-controlled cases. We will further quantify the sensitivity of clinically relevant accumulated-dose metrics to registration uncertainty and assess whether DIR-based accumulated-dose metrics improve outcome prediction compared with metrics derived from rigid dose accumulation.

\begin{credits}
\subsubsection{\ackname} This study was funded by NIH R44CA183390.

\subsubsection{\discintname}
The authors have no competing interests to declare that are
relevant to the content of this article.
\end{credits}

%
%
%
%
\bibliographystyle{splncs04}
\bibliography{references_miart_v2}
\end{document}